# Photoinjector design for the LCLS[*]

P.R. Bolton[a], J.E. Clendenin[a], D.H. Dowell[a], M. Ferrario[b], A.S. Fisher[a],
S.M. Gierman[a], R.E. Kirby[a], P. Krejcik[a], C.G. Limborg[a], G.A. Mulhollan[a],
D. Nguyen[c], D.T. Palmer[a], J.B. Rosenzweig[d], J.F. Schmerge[a], L. Serafini[e], X.-J. Wang[f]

[a]Stanford Linear Accelerator Center, Stanford, CA 94309, USA
[b]INFN-LNF, 00044 Frascati (Roma), IT
[c]LANL, Los Alamos, NM 87545, USA
[d]UCLA, Los Angeles, CA 90095, USA
[e]INFN-MI, 20133 Milano, IT
[f]BNL, Upton, NY 11973, USA

Corresponding author: J.E. Clendenin, P.O. Box 4349, Stanford, CA 94309, USA. Tel.: +650-926-2962; fax: +650-926-8533. E-mail address: clen@slac.stanford.edu.

Abstract

The design of the Linac Coherent Light Source assumes that a low-emittance, 1-nC, 10-ps beam will be available for injection into the 15-GeV linac. The proposed rf photocathode injector that will meet this requirement is based on a 1.6-cell S-band rf gun equipped with an emittance compensating solenoid. The booster accelerator with a gradient of 25 MV/m is positioned at the beam waist coinciding with the first emittance maximum, i.e., the "new working point." The uv pulses required for cathode excitation will be generated by tripling the output of a Ti:sapphire laser system. Details of the design and the supporting simulations are presented.

Contributed to
The 23[rd] International Free Electron Laser Conference
Darmstadt, Germany
20-24 August 2001

---

[*] Work supported by Department of Energy contract DE-AC03-76SF00515.

## 1. Introduction

The proposed Linac Coherent Light Source (LCLS) is an x-ray free electron laser (FEL) that will use the final third of the SLAC 3-km linac for the drive beam. The performance of the LCLS in the 1.5 Å regime is predicated on the availability of a 1-nC, 100-A beam at the 150-MeV point with normalised rms transverse emittance of 1 µm. The initial injector design made use of a low-gradient booster as was then the standard [1]. The calculated emittance using the initial design was close to the LCLS requirement, but only if the halo particles were ignored. A principal goal of the redesign was to reduce the calculated emittance to provide "headroom" for a practical machine. In addition, since the existing shielded vault at the SLAC linac is too small for the earlier design, a second goal was to decrease the physical length of the injector.

The basic layout of the injector is shown in Fig. 1. This layout is consistent with the concept of a "new working point" that was introduced in 1999 [2]. The basic difference seen in this design compared to the earlier version is that the drift distance between the gun and booster is greater and also the booster accelerating gradient is much higher. Note that a relatively weak solenoidal field is now used around the first booster section instead of discrete focusing coils between sections.

## 2. Gun, cathode, and laser

The LCLS gun will be a modified version of the BNL/SLAC/UCLA symmetrized 1.6-cell rf photocathode gun [3]. The principal differences are improved cooling [4] to allow operation at 120 Hz and the addition of a load lock. A load lock provides a reliable way to quickly change a cathode under controlled environmental conditions without breaking the gun vacuum that would require a rebake of the gun. In addition, by preparing the cathode in a separate ultra-clean facility and transporting it to the gun under vacuum, the quantum efficiency (QE) value and uniformity can be optimised.

A copper photocathode is chosen to permit the entire end plate of the half cell to be formed from a single piece, allowing gun operation at the highest possible field values. The QE for Cu illuminated with uv light depends on surface preparation, but $10^{-5}$ for normal incidence at 266 nm in a non-load-locked gun is achievable [5,6,7]. Much better QE is available from copper installed through a load lock as illustrated by the data of Fig. 2.

Photocathodes made from single-crystal Cu have proven to have not a only high QE but also low dark current [8]. In addition, the QE uniformity appears to be superior to that of polycrystalline Cu. Single-crystal copper boules larger than the cell end plate are available..

The changes in the laser system described in reference [1] are relatively minor. To increase stability, the Ti:sapphire oscillator is now pumped by a highly stable Nd:YV0$_4$ laser. Also the oscillator pulse repetition rate is now 79.33 MHz to allow the timing of the laser pulses to be locked to the phase of the 2856 MHz rf driving the gun and linac. As a reminder, the two-stage four-pass bow-tie Ti:sapphire amplifier and associated optics delivers to the cathode a 500 μJ highly stable pulse at 266 nm with uniform spatial and temporal shapes at a rate of 120 Hz. This energy is sufficient to produce the desired charge if the QE is at least $10^{-5}$.

## 3. Simulations

Using the semi-analytic code HOMDYN [9], a wide range of injector parameters was investigated earlier under the constraints imposed by the invariant-envelope [10] matching condition: injection into the matched accelerating gradient of the booster at a laminar waist. As a result, it was found that by adjusting the gun solenoid strength so that the waist also occurs when the emittance has its relative maximum, the second emittance maximum can be shifted to higher energy with a lower final emittance value than previously achieved [2,11]. This new configuration is here referred to as the "new working point."

The LCLS injector has been designed using version 3 of the LANL-maintained code PARMELA to establish the details of the "new working point." Both the spatial and temporal charge distributions at the cathode were assumed to be uniform. A normalized rms thermal emittance of 0.26 µm was used [12].

Using only the gun, solenoid, and the immediately following drift space, the first emittance minimum after the solenoid was optimized by varying the gun solenoidal field and the beam radius at the cathode. A value of $B_z$ = 3.15 kG and hard-edge radius of 1 mm was found to be optimum. The emittance minimum very nearly coincides with the "new working point." A slightly larger value of $B_Z$ is found here than with HOMDYN, consistent with the solenoid being displaced somewhat downstream because of the physical interference of the gun structure.

Next the position of the booster with an accelerating gradient of 25 MV/m was optimized followed by the position and field of the linac solenoid. The injector parameters determined in this manner are summarized in Table 1. The cathode to end-of-booster distance is now only 7.9 m, which allows the injector to fit in the existing slightly-modified shielded vault.

For the parameters of Table 1, the normalized rms emittance, $\varepsilon_n=\gamma\varepsilon$, for a risetime of 0.5 ps and the rms beam size, $\sigma$, are shown in Fig. 3 as a function of distance from the cathode. At the end of the booster, $\varepsilon_n$ is now 0.8 µm compared to 1.1 for the core charge in the earlier design [13]. The normalized transverse phase space at the exit of the booster for 100K particles is shown in Fig. 4. The upper right plot is the normalized phase space where $x_n$ ($x'_n$) is the rms beam size normalized such that an emittance ellipse of 1 µm can be represented by a circle of dimensionless-radius unity; i.e., $x_n = 10^3 \sigma_x \sqrt{\gamma/\beta_T}$ and $x'_n = 10^3 \sigma_{x'} \sqrt{\gamma\beta_T}$, where $\beta_T$ is the Twiss parameter for the beam size at the location of interest. At the exit of the booster where $\gamma=300$, $\beta_x=\beta_y\sim 11$ m. The upper left plot is a normalized scatter plot. The charge distribution projected onto the normalized x-axis is shown in the lower right. In the lower left, the normalized rms slice emittance in $x$ and $y$, as a function of axial distance along the bunch, is shown. The projected value is given by the horizontal dashed line.

The same PARMELA simulations yield an integrated (slice) energy spread at the booster exit that is within 0.18 (0.005)%.

## 4. Conclusions

An rf photoinjector for the LCLS based on the "new working point" configuration has been described. The injector will utilize a 1.6-cell rf gun designed for 120 Hz and equipped with a Cu photocathode and load lock. The cathode is illuminated with 500 µJ of uv light provided by a frequency-tripled Ti:sapphire laser. At the output of the injector, the 150 MeV beam has a transverse emittance that is now calculated to be well below the 1-µm requirement and the slice emittance is about 25% (relative) lower except for the head and tail. The integrated (slice) energy spread is also very low.

# References

[1] R. Alley et al., Nucl. Instrum. and Meth. A 429 (1999) 324.
[2] M. Ferrario et al., in The Physics of High Brightness Beams, eds. J. Rosenzweig, L. Serafini, World Scientific (2000), p. 534.
[3] D.T. Palmer et al., SPIE 2522 (1995) 514.
[4] X.J. Wang et al., "High-Rep Rate Photocathode Injector for LCLS," contributed to the 2001 Particle Accelerator Conference, June 18-22, 2001, Chicago, IL.
[5] T. Srinivasan-Rao et al., J. Appl. Phys. 69 (1991) 3291.
[6] P. Davis et al., in Proc. of the 1993 Particle Accelerator Conference, p. 2976.
[7] E. Chevallay et al., Nucl. Instrum. and Meth. A 340 (1994) 146.
[8] P.R. Bolton et al., "Transverse emittance measurements on an S-band photocathode rf electron gun," this conference. See also D.T. Palmer et al., in The Physics of High Brightness Beams, eds. J. Rosenzweig, L. Serafini, World Scientific (2000), p. 439.
[9] M. Ferrario et al., Part. Acc. 52 (1996).
[10] L. Serafini, J. Rosenzweig, Phys. Rev. E 55 (1997) 7565.
[11] M. Ferrario et al., Proc. of the 7th European Particle Accelerator Conference, Vienna (2000) 1642.
[12] J.E. Clendenin et al., in The Physics of High Brightness Beams, eds. J. Rosenzweig, L. Serafini, World Scientific (2000), p. 504. Recent measurements using a very low charge beam predict $\varepsilon_{n,rms} \sim 0.5$ μm for this case: see W.S. Graves et al., "Measurement of Thermal Emittance for a Copper Photocathode," in Proceedings of the 19[th] Particle Accelerator Conference, Chicago, 2001 (to be published).
[13] A significant fraction of this improvement results from using a uniform temporal distribution rather than the previously required truncated Gaussian.

Table 1
Parameters and PARMELA results

*For each bunch*
Charge at cathode/at booster exit             1.0/1.0 nC
Shape at cathode spatial/temporal             Uniform/Uniform
Radius at cathode, hard-edge                  1.0 mm
Length at cathode/at booster exit             10/10 ps FWHM

*For assumed peak rf field in gun of 140 MV/m*
Extraction phase                              32°
Gun solenoid axial field                      3.15 kG
Cathode-to-booster entrance (exit)            1.4 (7.9) m
L0 gradient (phase L01/L02)                   25 MV/m (-2.5/+2.6°)
Linac solenoid axial field/length             -1.75 kG/0.9 m
Energy at booster exit                        150 MeV
Integrated (slice) energy spread              0.18 (0.005)% rms
$\varepsilon_{n,rms}$ for 0.5 (1.0) ps rise time,
    $\varepsilon_{n,th}$ included, 100K particles    0.80 (0.95) μm

**Figure Captions**

Fig. 1. Schematic layout (not to scale) showing only the principal beamline elements, the location of the diagnostics, and the rf distribution system. In the figure are shown the rf gun (G), the emittance compensating solenoid (S1), charge coupled devices (CCD), klystrons (K), the focusing solenoid (S2) around the first 3-m accelerating section (L0-1) of the booster.

Fig. 2. QE of copper as a function of quantum energy measured with low (22 V) dc bias with the surface untreated after installation in the analysis system using a load lock.

Fig. 3. Transverse normalized rms emittance as a function of distance from the cathode for 100K particles. A rise time of 0.5 ps is assumed. A normalized rms thermal emittance of 0.26 µm is included.

Fig. 4. Normalized transverse phase space out of the booster for 100K particles. Upper are scatter (left) and phase space (right) plots using normalized dimensions such that a normalized rms emittance of 1 µm can be represented by a circle of dimensionless-radius unity as drawn in the right plot. Lower plots are transverse normalized rms slice emittances (left) along the bunch z-axis (bunch head at positive Δz) and the projection of charge (right) along the $x_n$-axis.

Figure 1

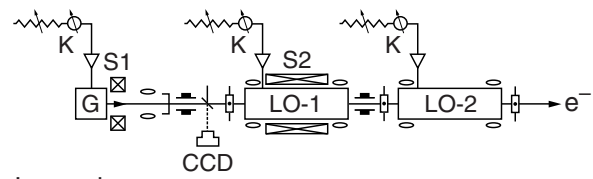

Legend

⚏ Solenoid   ⊥ Toroid   ] Faraday Cup
⊰ x and y Corrector   ⌄ Viewing Screen   ▯ BPM

10-2001
8612A2

**Figure 2**

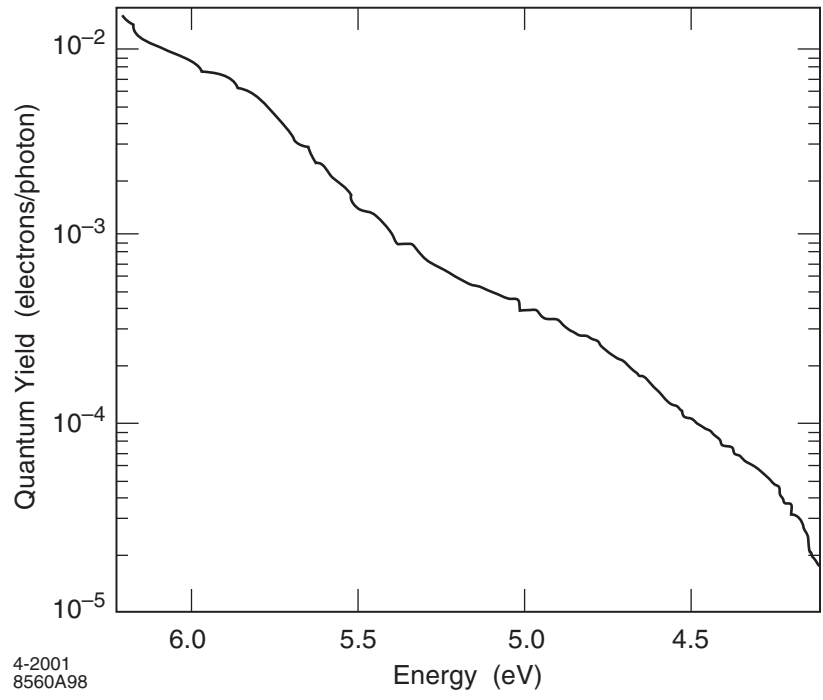

**Figure 3**

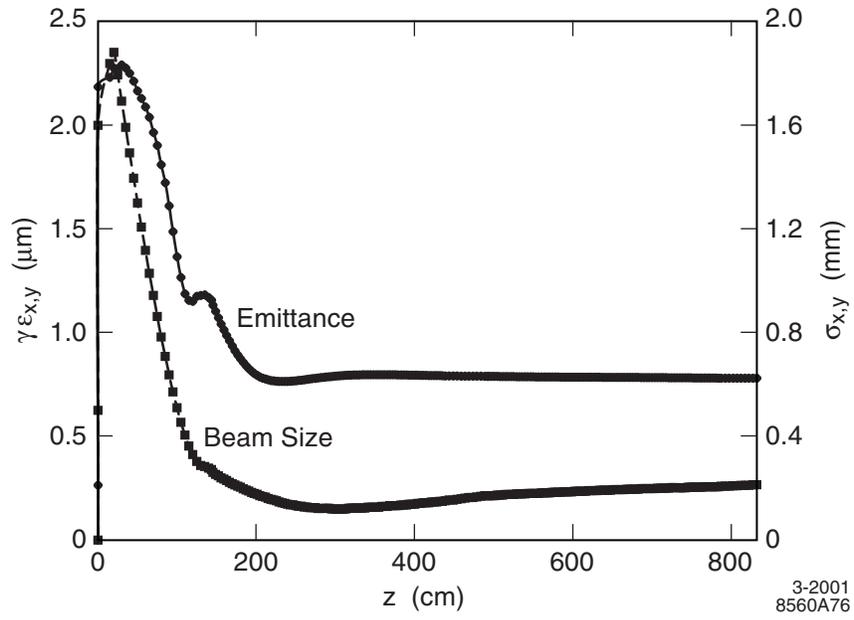

**Figure 4**

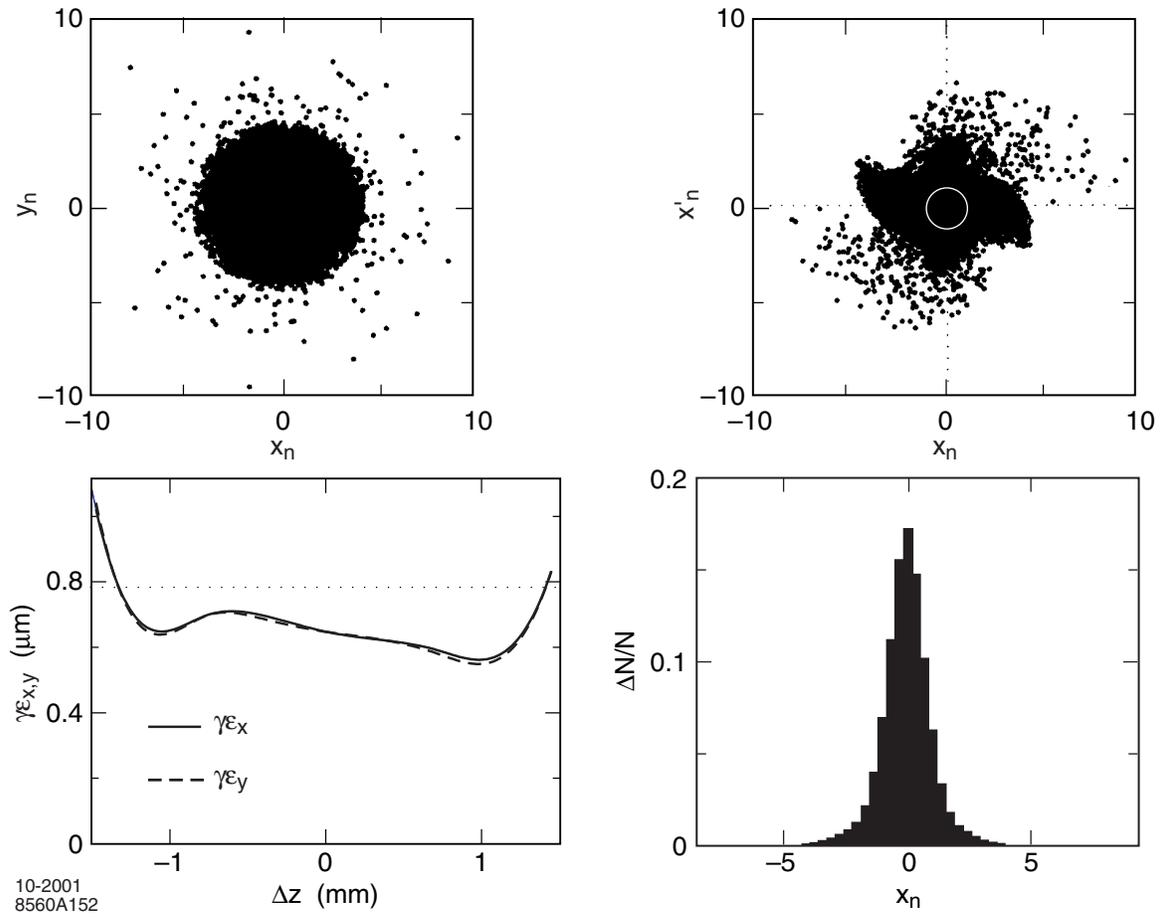